\documentclass[aps,prd,twocolumn,showpacs,floatfix,nofootinbib,superscriptaddress]{revtex4-1}
\usepackage{amsmath}
\usepackage{amssymb}
\usepackage{graphicx}
\usepackage{hyperref}
\usepackage{color}
\usepackage{mathtools}
\usepackage{times}

\newcommand{\DDD}{D}          
\newcommand{\bDD}{\bar\DDD}  
\newcommand{\dD}{\de\DDD}     
\newcommand{\dd}{{\cal\DDD}}  
\newcommand{\Ddd}{\de\dd}
\newcommand{\SSS}{\mathcal{S}} 
\newcommand{\FFF}{F}           
\newcommand{\IS}{\tilde{\SSS}}  
\newcommand{\zm}{z_\up{m}}     
 
\newcommand{\hdd}{\hat{\bar\dd}}
\newcommand{\GD}{\mathcal{G}}
\newcommand{\TT}{\mathcal{T}}
\newcommand{\TTT}{\mathcal{M}}
\newcommand{\up}[1]{{\rm #1}}
\newcommand{\bdv}[1]{{\bf #1}}
\newcommand{\beeq}{\begin{equation}}
\newcommand{\eneq}{\end{equation}}
\newcommand{\bear}{\begin{eqnarray}}
\newcommand{\enar}{\end{eqnarray}}
\newcommand{\nnn}{\nonumber \\}

\newcommand{\AVE}[1]{\left\langle#1\right\rangle}
\newcommand{\RA}{\rightarrow}
\newcommand{\pa}{\partial}
\newcommand{\OO}{\mathcal{O}}
\newcommand{\De}{\Delta}
\newcommand{\ga}{\gamma}
\newcommand{\de}{\delta}
\newcommand{\HH}{\mathcal{H}}   
\newcommand{\rbar}{\bar r}       
\newcommand{\mpc}{{\rm Mpc}}
\newcommand{\hmpc}{{h^{-1}\mpc}}

\begin{document}

\title{Maximum Cosmological Information from Type-Ia Supernova Observations}

\author{Jaiyul Yoo}
\email{jyoo@physik.uzh.ch}
\affiliation{Center for Theoretical Astrophysics and Cosmology,
Institute for Computational Science,
University of Z\"urich, Winterthurerstrasse 190,
CH-8057, Z\"urich, Switzerland}
\affiliation{Physics Institute, University of Z\"urich,
Winterthurerstrasse 190, CH-8057, Z\"urich, Switzerland}

\date{\today}

\begin{abstract}
Type-Ia supernova observations yield estimates of the luminosity distance,
which includes not only the background luminosity distance, but also
the fluctuation due to inhomogeneities in the Universe. These fluctuations
are spatially correlated, hence limiting the cosmological information.
In particular, the
spatial correlation of the supernova
 host galaxies is a dominant source of the 
fluctuation in the luminosity distance measurements. Utilizing 
the recent theoretical framework that accurately quantifies the information 
contents accounting for the three-dimensional
correlation of the observables on the past-light cone,
we compute the maximum cosmological information
obtainable from idealized supernova surveys, where an infinite number of
observations are made over the full sky without any systematic errors up to
a maximum redshift~$\zm$.
Here we consider two cosmological
parameters $\Omega_m$ and $w_0$ and show that the cosmological information 
contents are a lot more reduced than previously computed in literature.
 We discuss how these fundamental 
limits set by cosmic variance can be overcome.
\end{abstract}

\maketitle

\section{Introduction} 
Measurements of type-Ia supernova provide a powerful
way to probe cosmology \cite{RIFIET98,PEADET99}. A great amount of efforts
in recent decades have been devoted to increase the survey volume and reduce 
the intrinsic systematic errors in the luminosity distance measurements
\cite{SDSSSN09,SNLS11,SNLS14,ESSENCE16,PSSN18}.  
The error budget in  supernova
surveys is composed of intrinsic statistical errors due to the variation
in the absolute luminosity of the individual type-Ia supernovae and
systematic errors associated with the light-curve calibration. Since the former
is statistical by nature, its contribution to the error budget can be reduced
by increasing the number of supernova observations. The latter is, however,
more difficult to quantify and control, as it depends on many uncertain
factors such as the physical mechanism of the type-Ia supernova explosion,
the environment of the host galaxy, and so on (see 
\cite{SDSSSN09,SNLS11,SNLS14,ESSENCE16,PSSN18} for details).
 In addition, it is well-known that 
supernova observations yield an estimate~$\DDD(z,\hat n)$
of the luminosity distance that includes not only
the background luminosity distance~$\bDD(z)$ at the observed redshift~$z$, 
but also the fluctuation~$\dD(z,\hat n)$ at the observed position
specified the angular direction~$\hat n$ and the redshift~$z$, 
where $\DDD:=\bDD(1+\dD)$.
The fluctuation in the luminosity distance arises, because
the observed flux, the redshift, and the angular position
are affected by the large-scale inhomogeneities 
in the Universe through the light propagation from the source to the observer.
Hence, all the measurements of the luminosity distance are spatially correlated
due to its fluctuations, and this correlation also contributes to the
error budget.

Since the pioneering work \cite{SASAK87}, many groups showed 
\cite{BODUGA06,HUGR06,CLELET12,KAHU15a,YOSC16,FLCLMA16,BIYO16,SCYO17}
that the
fluctuation~$\dD$ in the luminosity distance contains the line-of-sight
peculiar motion~$V$ of the host galaxy, the gravitational potential 
contribution~$\phi$, and the gravitational lensing effect~$\kappa$. 
The peculiar velocity is the dominant source of correlation at low redshift,
while the gravitational lensing effect takes over at higher redshift.
The effects of these inhomogeneities have been investigated 
\cite{HUGR06,BODUKU06,CLELET12,BEGAET12a,BEGAET13a,BEDUET14} in the past.
Furthermore, since the progenitors of supernovae are associated with galaxies,
supernova observations are also {\it biased} \cite{KAHU15a,FLCLMA16},
as their host galaxies are biased against the underlying matter distribution,
and this fluctuation of the observed host galaxies 
(defined as~$\de_g$ in Eq.~[\ref{dobs}] below) 
indeed constitutes the {\it dominant}
contribution to the correlation in the supernova observations.

Therefore, in deriving the cosmological constraints from supernova surveys,
it is important to take into consideration all the correlation of the 
luminosity distance measurements. However, previous analysis often ignored 
the {\it radial correlation} or the correlation due to the {\it 
host galaxies}.
Accounting for all the effects described above, a complete theoretical
formalism was derived in 
Ref.~\cite{YOMIET19} for accurately quantifying the cosmological information 
contents on the light cone under the assumption that the fluctuations are 
at the linear order and Gaussian-distributed.
In this paper, we apply this formalism to supernova surveys
in the 3D light cone volume and
compute the maximum cosmological information
contents that are available to us in the idealized  surveys, where 
an infinite number of supernova observations are made without any systematic
errors in full sky over all redshift up to a given maximum
redshift~$\zm$.

\section{Observed Data Set} 
Individual observations of type-Ia supernovae
yield an estimate of the luminosity distance~$\DDD(z,\hat n)$ 
at the observed redshift~$z$ and angular direction~$\hat n$, but the observed
data set~$\dd(z,\hat n)$ altogether are described 
as the luminosity distance~$\DDD$ weighted by the number~$dN_g$ of the
observed host galaxies in a given volume~$d\bar V$ determined by the redshift 
bin~$dz$ and the angular bin~$d^2\hat n$ \cite{YOMIET19}:
\beeq
\label{obsdata}
\dd(z,\hat n)=\DDD(z,\hat n){dN_g(z,\hat n)\over N_g^\up{tot}}
:=\bar\dd(z)\left[1+\Ddd(z,\hat n)\right]~,
\eneq
where we defined the background~$\bar\dd(z)$ and
the (dimensionless) fluctuation~$\Ddd(z,\hat n)$ around it.
 In the case of one supernova
observation, the total number of the observed host galaxies
$N_g^\up{tot}:=\sum dN_g$ is unity,
and the observed data set is just an estimate of the luminosity distance 
$\dd(z,\hat n)=\DDD(z,\hat n)$. With more observations in the data set,
more weight is naturally given to the estimates~$D(z,\hat n)$ in an over-dense
region, where more supernova events are observed. 

It is important to note that our theoretical description~$\dd$ in 
Eq.~\eqref{obsdata} is {\it not}
the luminosity distance itself at a given position, but a description 
of the (discrete) observed data set in the limit $N_g^\up{tot}=\infty$.
In practice, there exist a finite number of observed data points
for the luminosity distance over the survey region, 
and the observers treat each individual point
equally. For example,
there will be no measurement or data point ($\dd=0$) in the observed data set
in a void, where there is no observed host galaxy ($dN_g=0$), even though
the luminosity distance to such void is non-zero ($\DDD\neq0$).
This aspect of the observed data set is correctly
described by Eq.~\eqref{obsdata} in the continuum limit. 

The number weight~$dN_g$ is the physical
number density~$n_p$ of the host galaxies times the physical volume~$dV_p$
described by the observed redshift bin~$dz$ and the angular bin~$d^2\hat n$:
\beeq
\label{dobs}
dN_g(z,\hat n)=n_gdV_p:={d\bar N_g\over dzd^2\hat n}(z)
~dzd^2\hat n\left[1+\de_g(z,\hat n)\right]~,
\eneq
where the background part is simply the redshift distribution of the host
galaxies in a homogeneous universe in terms of the Hubble parameter~$H$
and the comoving angular diameter distance~$\rbar$
\beeq
\label{dNdz}
{d\bar N_g\over dzd^2\hat n}(z):={\bar n_g(z)\rbar^2(z)\over H(z)(1+z)^3}~.
\eneq
Defined as above, 
the fluctuation~$\de_g(z,\hat n)$ of the observed host galaxies is shown to be
gauge-invariant \cite{YOFIZA09,YOO10,BODU11,CHLE11,JESCHI12,YOO14a},
and it includes the source fluctuation (such as the galaxy bias and the 
magnification bias) and the volume fluctuation (such as the redshift-space
distortion and the relativistic effects) \cite{YOO09}. 

Therefore, the background~$\bar\dd$ and the fluctuation~$\Ddd$ 
of the observed data set are \cite{YOMIET19}
\bear
\label{bDD}
&&\bar\dd(z)={\bDD(z)\over\bar N_g^\up{tot}}
{d\bar N_g\over dzd^2\hat n}(z)~dz d^2\hat n =:\hdd(z)~dzd^2\hat n~,\\
\label{Ddd}
&&1+\Ddd(z,\hat n)={\left[1+\dD(z,\hat n)\right]\left[1+\de_g(z,\hat n)\right]
\over1+\de N_g^\up{tot}}~,
\enar
where the total number~$N_g^\up{tot}=\bar N_g^\up{tot}(1+\de N_g^\up{tot})$
of the observed host galaxies is split into two (dimensionless)
constants for later convenience
\bear
\label{Ntot}
\bar N_g^\up{tot}&:=&4\pi\int_0^{\zm} dz~{d\bar N_g\over dzd^2\hat n}
(z)~,\\
\label{dNtot}
\de N_g^\up{tot}&:=&4\pi\int_0^{\zm} dz~{1\over\bar N_g^\up{tot}}
{d\bar N_g\over dzd^2\hat n}(z)\int {d^2\hat n\over4\pi} ~\de_g(z,\hat n)~.
\enar

\section{Single Redshift Bin} 
We first consider supernova observations 
at a given redshift~$z_*$ with small redshift bin~$\De z$ 
to illustrate how individual fluctuations
affect the luminosity distance estimate. The total number weight
of the observed host galaxies in this case is
\beeq
\bar N_g^\up{tot}=4\pi\De z{d\bar N_g\over dzd^2\hat n}~,\qquad
\de N_g^\up{tot}=\int{d^2\hat n\over4\pi}~\de_g(\hat n)~,
\eneq
directly related to the redshift distribution at the given redshift~$z_*$, and
the observed data set is then
\beeq
\label{2dobs}
\dd(\hat n)=\bDD(z_*){d^2\hat n\over4\pi}\left[1+\Ddd(\hat n)\right]~,
\eneq
where we suppressed the dependence on the survey redshift~$z_*$
and the fluctuation at the linear order in perturbations is
\beeq
\label{deltasing}
\Ddd(\hat n)\simeq\dD(\hat n)+\de_g(\hat n)-\de N_g^\up{tot}+\OO(2)~.
\eneq
The number weight drops in the background part to yield $\bar\dd=\bDD$
up to numerical factors, as all the observed data at the same redshift
are summed up. The fluctuation~$\Ddd$ is, however, dependent upon
the fluctuation~$\de_g$ of the host galaxies as well as the fluctuation~$\dD$
in the luminosity distance. When averaged over the observed data set or
the angle at~$z_\star$,
Eq.~\eqref{deltasing} shows that~$\de_g$ and~$\de N_g^\up{tot}$ drop out 
\beeq
\Ddd_0:=\int{d^2\hat n\over4\pi}~\Ddd(\hat n)=\int{d^2\hat n\over4\pi}
~\dD(\hat n)=:\dD_0~,
\eneq
though this is valid only at the linear order.

The fluctuation~$\dD$ in the luminosity distance at the linear order 
is often computed in the conformal Newtonian gauge 
\cite{BODUGA06,HUGR06,KAHU15a,YOSC16}
\beeq
\label{single}
\dD(\hat n)\approx V_s-{1+z_*\over H\rbar}\left(V_s-V_o\right)-\kappa~,
\eneq
where $V_s$ and~$V_o$ are the line-of-sight peculiar velocity at the source
and the observer positions and $\kappa$~is the lensing convergence.
Since the gravitational potential contribution
to the luminosity distance is about a percent level correction to 
Eq.~\eqref{single} (see Figure~3 in \cite{BIYO16}),
we ignored the gravitational potential contributions to~$\dD$. 
As emphasized
\cite{BIYO16,SCYOBI18},
the individual components in~$\dD$ are gauge-dependent
and {\it not} associated with any physical quantity, while
the full expression for~$\dD$ is gauge-invariant. For example, $V_o$~should
{\it not} be linked to the velocity we measure from the CMB dipole,
because the former has different values in other gauge conditions,
while the value of the latter is uniquely fixed by observations, independent
of our gauge choice. Hence,  one {\it cannot}
simply remove the peculiar velocity~$V_o$ at the observer position in computing
the luminosity distance, and it was shown \cite{BIYO16} that the peculiar
velocity~$V_o$ at the observer position is a dominant contribution
$V_o>V_s$ and this procedure of arbitrarily removing perturbation
contributions at the observer position was the source of the infrared
divergences in previous calculations \cite{BAMARI05,KOMAET05a}.

On a single redshift bin, the correlation of the luminosity distance
fluctuation is conveniently decomposed in terms of angular power 
spectrum~$C_l$,
\beeq
\xi_{ij}:=\AVE{\Ddd(\hat n_i)\Ddd(\hat n_j)}:=
\sum_l{2l+1\over4\pi}C_l P_l(\ga_{ij})~,
\eneq
where $\gamma_{ij}:=\hat n_i\cdot\hat n_j$ and $P_l(x)$ is the Legendre
Polynomial. 
Under the assumption of Gaussianity, the information contents of the observed
data set for a given
set of cosmological parameters~$p_\mu$ are quantified in Ref.~\cite{YOMIET19}
by using the Fisher matrix 
\beeq
\label{2dFisher}
\FFF_{\mu\nu}={4\pi\over C_0}\left({\pa\ln\bDD(z_*)\over\pa p_\mu}\right)
\left({\pa\ln\bDD(z_*)\over\pa p_\nu}\right)+\cdots~,
\eneq
where we ignored the cosmological information contained in the correlation 
$\xi_{ij}(p_\mu)$ shown as dotted in the equation
and the monopole power~$C_0$
is solely determined by the peculiar velocity at the source position
\beeq
\label{angavg}
\Ddd_0=\dD_0=\left(1-{1+z_*\over H\rbar}\right)\int{d^2\hat n\over4\pi}~V_s
(\hat n)=:{a_0\over\sqrt{4\pi}}~.
\eneq
The explicit expression for the monopole power $C_0=\AVE{|a_0|^2}$  is 
\beeq
C_0=\left[{H f\over1+z_*}\left(1-{1+z_*\over H\rbar}\right)\right]^2{2\over\pi}
\int dk~P_\varphi(k)\TT^2_m(k)j_0^{\prime2}(k\rbar)~,
\eneq
where $f$ is the logarithmic growth rate, 
$\Delta^2_\varphi(k):=k^3P_\varphi/2\pi^2$ is the dimensionless scalar
power spectrum, $\TT_m(k)$
is the matter transfer function at~$z_*$, and $j_0(x)$ is the spherical
Bessel function. With $j'_0(x)=-j_1(x)$, the monopole power~$C_0$ is regular
in the limit $z_*\RA0$.

The Fisher matrix expression shows that 
(1)~individual observations on a single redshift bin essentially yield 
the average over the angle and the ensemble average of the estimate is 
the background luminosity distance~$\bDD(z_*)$, (2)~the information contents 
depend on the sensitivity of~$\bDD$ to the cosmological parameters~$p_\mu$,
and
(3)~upon average over the sky, the estimate is limited by the cosmic variance 
of the monopole power~$C_0$. In~$C_0$ (or $a_0$),
the host fluctuation, the lensing effect, and the velocity at the observer
position completely drop out. While the correlation~$\xi_{ij}$ is
a function of cosmology \cite{BIYO17},
the correlation of the luminosity distance in practice
is {\it rarely} measured, and 
we ignore this contribution to the cosmological information
(see, however, \cite{BIYO17}).

\section{3D Light-Cone Volume} 
Supernova observations are not confined
to a single redshift bin; naturally, they cover a range of redshifts, 
increasing
the leverage to constrain cosmology. Since measurements of the luminosity
distance at different redshift bins are also correlated, the increase in the 
cosmological constraining power is somewhat limited. 
A critical difference in a 3D light-cone volume is that the number weight
of the host galaxies changes in redshift, primarily due to the volume, but 
also due to the evolution of the host galaxies. Here we assume
that the physical
number density of the host galaxies evolves as $\bar n_g(z):=n_0(1+z)^\alpha$,
where $\alpha=3$ represent a constant comoving number density such as 
dark matter. Compared to Eq.~\eqref{2dobs}, the expectation~$\bar\dd(z)$ 
of the observed data set in Eq.~\eqref{bDD} has the number weight factor, 
while the overall normalization constant~$n_0$ is irrelevant in our discussion.

Similarly, the number weight plays a role at the perturbation level~$\Ddd$. 
The
fluctuation~$\de_g$ of the host galaxies is modeled here as the sum of the
redshift-space distortion~$\de_z$ \cite{KAISE87} and the gravitational
lensing effect~$\kappa$: $\de_g(z,\hat n)\approx \de_z-2\kappa$, where
we ignored the velocity and the gravitational potential 
contributions in the full relativistic expression \cite{YOO14a}, as 
the density fluctuation is the dominant contribution and the lensing effect
(derivatives of the gravitational potential) is comparable. 
For the fluctuation~$\dD$
in the luminosity distance, we again ignore the velocity and the gravitational
potential contributions to be consistent with the approximation for~$\de_g$:
$\dD(z,\hat n)\approx-\kappa$.
Therefore, the fluctuation in the observed data set in Eq.~\eqref{Ddd} is
$\Ddd(z,\hat n)\approx\de_z-3\kappa-\de N_g^\up{tot}$,
and its average over the sky at a given redshift is
\beeq
\label{mono3d}
\Ddd_0(z)\approx\int{d^2\hat n\over4\pi}~\de_z(z,\hat n)-\de N_g^\up{tot}(z)~,
\eneq
where the lensing contribution again vanishes upon average. Provided that
the luminosity function of the host galaxies is well-approximated by a simple
power law $\propto L^{-s}$ with the slope~$s$, there exists an extra
contribution~$(2s-1)\kappa$ in the expression~$\de_g$ due to the magnification
bias \cite{YOO14a}, but this contribution again vanishes upon average.

In a 3D light-cone volume, any observable quantities can be decomposed
in terms of spherical harmonics for the angular dependence and spherical
Bessel for the radial dependence \cite{YODE13,YOMIET19},
and the correlation of the luminosity 
distance fluctuation is then expressed by the spherical power spectrum
$\SSS_l(k,k')$,
\bear
\xi_{ij}&=&\AVE{\Ddd(z_i,\hat n_i)\Ddd(z_j,\hat n_j)}:=
4\pi\sum_{lm}\int dk\int dk'        \nnn
&&\times ~{kk'\over2\pi^2}\SSS_l(k,k') 
j_l(k\rbar_i)j_l(k'\rbar_j)Y_{lm}(\hat n_i)Y_{lm}^*(\hat n_j)~.\quad
\enar
The spherical power spectrum is a generalization of the standard (flat-sky)
power spectrum~$P(k)$, and they are equivalent, if~$P(k)$
is {\it isotropic} and there is {\it no} evolution along the
radial direction, i.e., $\SSS_l(k,k')=\de^D(k-k')P(k)$.
Under the assumption that the fluctuations are Gaussian distributed, the
information contents in a 3D light-cone volume can be quantified in terms
of the Fisher matrix \cite{YOMIET19} as
\beeq
\label{Fisher3d}
\FFF_{\mu\nu}=(4\pi)^2\int dk\int dk'~{kk'\over2\pi^2}
\IS_0(k,k')\GD_\mu(k)\GD_\nu(k') +\cdots,
\eneq
where $\IS_l(k,k')$ is the inverse spherical power spectrum, 
the Fourier kernel~$\GD_\mu(k)$ depends on cosmological parameters~$p_\mu$
\beeq
\label{GDk}
\GD_\mu(k):=\int_0^{\zm}dz~j_0(k\rbar_z){\pa\over\pa p_\mu}\ln\hdd(z)~,
\eneq
and we again ignored the cosmological information contained in the 
correlation $\xi_{ij}(p_\mu)$ shown as dotted.

Compared to the case for a single redshift bin,
the Fisher matrix for a 3D light-cone volume reveals that (1)~individual 
observations over the redshift range essentially yield~$\hdd(z)$ [defined
in Eq.~\eqref{bDD}], the background luminosity distance weighted by the 
redshift distribution of the host galaxies, including the volume factor,
(2)~the information contents depend on the sensitivity of the full background
quantity~$\hdd(z)$ to the cosmological parameters~$p_\mu$, {\it not}
just the background luminosity distance~$\bDD(z)$,  and (3)~the estimate
of~$\hdd(z)$ is limited by the cosmic variance described by the inverse
monopole power spectrum~$\IS_0(k,k')$, defined in Ref.~\cite{YOMIET19} as
\beeq
\label{inverse}
\bdv{\tilde S}_l=\left(\bdv{F}_l\bdv{S}_l\bdv{F}_l\right)^{-1}~,
\eneq
in the matrix notation, and the angular Fourier kernel is
\beeq
\bdv{F}_l(k,k'):={2k'\over\pi}\int_0^{\zm} dz~j_l(k\rbar_z)j_l(k'\rbar_z)~
\eneq
(see \cite{YOMIET19} for detailed derivations). For a single redshift bin,
the inverse power spectrum in the Fisher matrix is {\it literally} the 
inverse of the angular power spectrum, i.e., $\tilde C_l=C_l^{-1}$.
However, for a 3D light-cone volume, the inverse power spectrum~$\IS_l(k,k')$
is {\it not} the inverse of the spherical power spectrum~$\SSS(k,k')$,
i.e., $\IS_l(k,k')\neq\SSS_l^{-1}(k,k')$,
as apparent in the matrix inversion relation in Eq.~\eqref{inverse}.
The reason is that the evolution along the light cone mixes different
Fourier modes and the survey volume is finite, rather than an infinite
hypersurface. 

\begin{figure}[t]
\centering
\includegraphics[width=0.5\textwidth]{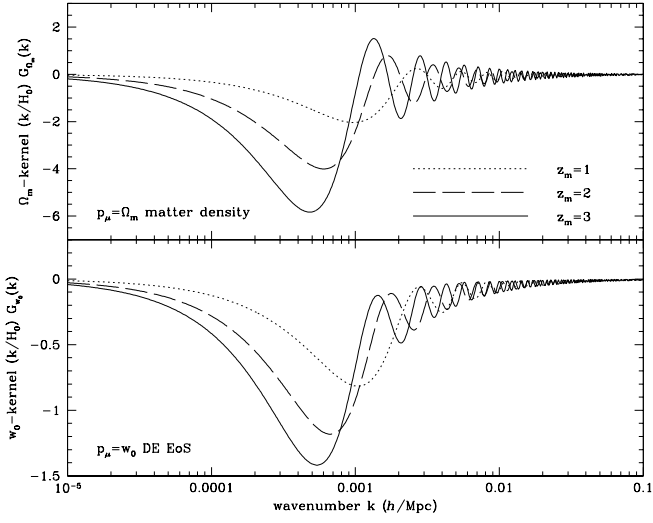}
\caption{Fourier kernel~$\GD_\mu(k)$ in Eq.~\eqref{GDk}, scaled with 
wavevector~$k$. The sensitivity of luminosity distance measurements
in a 3D light-cone volume to cosmological parameters is represented 
by $\GD_\mu(k)$ 
in the Fisher matrix in Eq.~\eqref{Fisher3d}. We consider 
all-sky supernova surveys with three different
maximum redshifts~$\zm$ denoted as solid ($\zm=1$), dashed ($\zm=2$), and
dotted ($\zm=1$).}
\label{fig:GDk}
\end{figure}

\begin{figure}[t]
\centering
\includegraphics[width=0.5\textwidth]{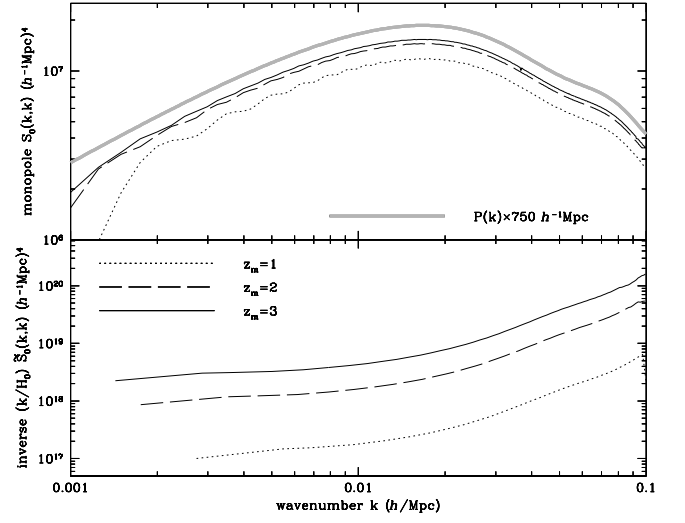}
\caption{Monopole $\SSS_0(k,k)$ and inverse monopole~$\IS_0(k,k)$ power
spectra for the surveys with three different maximum redshifts
shown in the legend.
Only the diagonal parts of the monopole spectra are plotted. {\it Upper:}
The monopole spherical power spectra~$\SSS_0(k,k)$ 
is plotted with three~$\zm$.
As a reference, the gray curve shows the matter
power spectrum~$P_m(k)$ at redshift zero, scaled with 750~$\hmpc$.
{\it Lower:} The inverse monopole power
spectra~$\IS_0(k,k)$ with three~$\zm$ is shown. $\IS_0(k,k)$
are scaled with wavenumber to highlight the contribution to the
Fisher matrix in Eq.~\eqref{FFapp}.}
\label{fig:S0}
\end{figure}

Using Eq.~\eqref{mono3d}, the monopole power spectrum can be computed as
\bear
\SSS_0(k,k')&=&4\pi\int d\ln\tilde k~\Delta_\varphi^2(\tilde k)
\bigg[\TTT_0^{\de_z}(k,\tilde k)-\TTT_0^{\de N_g^\up{tot}}\!\!\!\!
(k,\tilde k)\bigg]\nnn
&&\qquad \times \bigg[\TTT_0^{\de_z}(k',\tilde k)-\TTT_0^{\de N_g^\up{tot}}
\!\!\!\!(k',\tilde k)\bigg]~,
\enar
where the Fourier kernels for the angle-averaged
redshift-space distortion and the fluctuation in the number weight are
\bear
\label{Tz}
&&
\TTT_0^{\de_z}(k,\tilde k):=k\sqrt{2\over\pi}\int_0^{\rbar_\up{m}}
\!\! d\rbar~\rbar^2j_0(k\rbar)
\TT_z(\tilde k;\rbar)~,\\
&&
\label{TN}
\TTT_0^{\de N_g^\up{tot}}\!\!\!(k,\tilde k):=
k\sqrt{2\over\pi}\int_0^{\rbar_\up{m}} d\rbar~\rbar^2j_0(k\rbar) \\
&&\qquad\qquad\qquad\times
4\pi\int_0^{\zm}\!\!
dz~{1\over\bar N_g^\up{tot}}{d\bar N_g\over dzd^2\hat n}~\TT_z(\tilde k;z)~,
\nnn
&&
\label{RSD}
\TT_z(k;\rbar):=\left[
\left(b+\frac13f\right)j_0(k\rbar)-\frac{2f}3j_2(k\rbar)\right]
\TT_m(k;\rbar)~,~~~~~~~
\enar
and $\rbar_\up{m}:=\rbar(\zm)$. The dependence of the redshift-space
distortion kernel~$\TT_z(k)$ on $l=0$ and~$l=2$ arises due to the dependence on
$\mu_k^2$ of $\delta_z(k)=(b+f\mu_k^2)\de_m(k)$, where $b$~is the galaxy bias
factor, $\mu_k:=\hat n\cdot\hat k$, and $f$~is the logarithmic growth rate. The
kernels~$\TTT_0(k,\tilde k)$ represent the contributions to the angle
average in Eq.~\eqref{mono3d}, and they are not symmetric in arguments.

\section{Numerical Computation} 
To be specific, we adopt the best-fit
$\Lambda$CDM model cosmological parameters presented in Table~7 (Planck alone)
of the {\it Planck} 2018 result \cite{PLANCKover18}. 
For simplicity, we first assume that 
the host galaxies are described by the matter distribution ($b=1$ and
$\alpha=3$), and we ignore the redshift-space distortion ($f\equiv0$).
Figure~\ref{fig:GDk} shows the Fourier
kernels~$\GD_\mu(k)$ for two cosmological parameters with three different
maximum redshifts~$\zm$ of idealized supernova surveys.
The kernel~$\GD_\mu(k)$ multiplied by a wavenumber contributes to 
the Fisher matrix $\FFF_{\mu\nu}$ in Eq.~\eqref{Fisher3d}, 
and the product is bounded at all~$k$, exhibiting the peak
contribution around the characteristic scale of the survey depth
$k\approx2/\rbar_\up{m}$. Compared to $\pa\ln\bDD/\pa p_\mu$ in
the single redshift bin,
the log-derivative in~$\GD_\mu(k)$ includes the volume factor, which enhances
the sensitivity by about factor two at a given redshift. The dependence of 
the evolution slope~$\alpha$ is rather weak, only through the total 
number~$\bar N_g^\up{tot}$ for~$\GD_\mu(k)$.

The upper panel of Figure~\ref{fig:S0} shows the spherical monopole power 
spectra~$\SSS_0(k,k)$ with three~$\zm$. The Fourier 
kernels~$\TTT_0(k,\tilde k)$ for $\SSS_0(k,\tilde k)$ behaves like 
a Dirac delta function due to two spherical Bessel functions in the kernels.
Consequently, $\SSS_0(k,\tilde k)$ is {\it nearly} diagonal. 
In the limit $k\RA\infty$, where the survey depth is comparatively large 
$\rbar_\up{m}\gg1/k$, the integrals
in Eqs.~\eqref{Tz} and~\eqref{TN} can be performed analytically by ignoring
the time-evolution of the transfer function~$\TT_m(k;\rbar)$ to yield
$\IS_0(k,k')\approx\de^D(k-k')b^2P_m(k)$,
where $P_m$ is the matter power spectrum averaged over the survey depth.
Since $\de^D(k-k)$ in $\SSS_0(k,k)$ is proportional 
to~$\rbar_\up{m}$, the monopole power $\SSS_0(k,k)$ with~$\zm=3$ is largest
among three, while the averaged matter power spectrum with~$\zm=3$ is lowest
among three. For a reference, the matter power spectrum today is plotted
as a gray curve, illustrating the similarity to $\SSS_0(k,k)$ at 
$k\rbar_\up{m}\gg1$. At low~$k$, the power in $\SSS_0(k,k)$ is reduced due to
the survey volume.

\begin{figure}[t]
\centering
\includegraphics[width=0.5\textwidth]{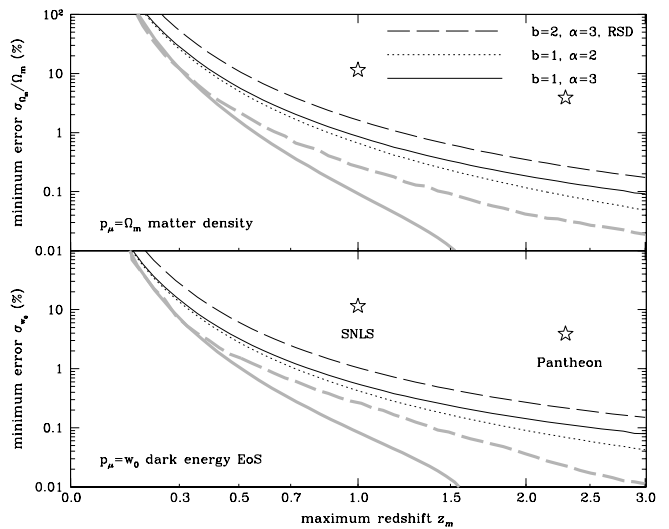}
\caption{Minimum fractional errors on two cosmological parameters 
from idealized supernova surveys alone
that measure infinite number of supernova events without
any systematic errors up to the maximum redshift~$\zm$. The correlation
of the fluctuations in the luminosity distance and the host galaxies
limits our ability to measure the cosmological parameters precisely
in the surveys. Various curves illustrate the dependence of
the model parameters associated with the host galaxies.
Gray curves show the change, if we ignore the radial correlation
(solid) or the host galaxy fluctuation (dashed). They represent the past
(imprecise) forecasts in literature.}
\label{fig:error}
\end{figure}

The lower panel in Figure~\ref{fig:S0} shows the inverse monopole power 
spectra~$\IS_0(k,k)$, scaled with a wavevector. Since~$\SSS_0(k,k')$ is nearly
diagonal, $\IS_0(k,k')$ in Eq.~\eqref{inverse} is nearly diagonal as well. 
However, the inverse spectrum~$\IS_0(k,k')$ is defined only for the wave
numbers that are approximately measurable in a given survey, i.e., 
$k\rbar_\up{m}\gtrsim1$. For any long wavelength modes ($k\rbar_\up{m}\ll1$),
the spherical power spectrum~$\SSS_0(k,k)$ has approximately the same
value close to zero, hence the matrix~$\bdv{S}_0$ is {\it not} invertible,
as the spherical Bessel functions vary little
for those wave numbers over the survey volume. 
So $\IS_0(k,k)$ is plotted only for $k\geq2\pi/\rbar_\up{m}$.
The inverse spherical power is very flat over the survey scales. 
Since $\IS_0(k,k')$
is nearly diagonal, the Fisher matrix in Eq.~\eqref{Fisher3d} can be 
re-arranged as
\beeq
\label{FFapp}
\FFF_{\mu\nu}\approx 8\int d\ln k~\left[\int dk'~k\IS_0(k,k')\right]
k\GD_\mu(k)~k\GD_\nu(k)~,
\eneq
illustrating that the cosmological information is proportional to the Fourier
kernel $k\GD_\mu(k)$ in Figure~\ref{fig:GDk} and the inverse power spectrum
$k\IS_0(k,k)$ in Figure~\ref{fig:S0}. For numerical computation 
of~$\FFF_{\mu\nu}$ in Eq.~\eqref{Fisher3d}, we compute the full 
matrices~$\bdv{S}_0$ and~$\bdv{\tilde S}_0$.

Figure~\ref{fig:error} illustrates the minimum fractional errors
on two cosmological parameters from idealized supernova surveys.
The minimum errors decrease with~$\zm$, as more volume is
included in the surveys. Compared to our fiducial model (solid), 
host galaxies with higher bias ($b=2$; dashed) are more correlated, 
reducing our leverage to constrain the cosmological parameters, and the
increase in the minimum errors are about the ratio of the (constant)
bias factors. As type-Ia supernovae occur in a highly biased region
(described by the bias factor in our model)  such
as massive galaxies or clusters of galaxies rather than field galaxies,
the observed data set from supernova surveys probes only the biased region
of the Universe, and its constraints deviate more from the cosmic mean.
The redshift-space distortion (RSD) also increases the
correlation of the host galaxies, though its impact is minor,
as the logarithmic growth rate~$f$ is small and two terms with~$f$
in Eq.~\eqref{RSD}
are partially cancelled. The host galaxies whose number density increases
in time ($\alpha=2$; dotted) effectively put more weight on the lower
redshift populations and increase $\delta N_g^\up{tot}(z)$,
reducing the fluctuation in the observed data set in Eq.~\eqref{mono3d}.
But its impact is small and visible only when the redshift depth is large.

Gray curves illustrate the change in the forecast, if the radial correlation
(solid) or the host galaxy correlation (dashed) is ignored. These cases
represent the current status in literature. In the former,
the survey volume is split into multiple redshift bins with width 
$\Delta z=0.05$, and the Fisher matrix in Eq.~\eqref{2dFisher} 
for each redshift bin are added, as if each redshift bin is independent.
This procedure of ignoring the radial correlation 
{\it significantly underestimates}
the cosmic variance and hence the minimum errors. In fact, the results
depend on the bin width~$\Delta z$ or the number of ``independent'' 
redshift bins. When the host galaxy fluctuation is ignored ($\de_g\equiv0$,
dashed gray), the observed data set Eq.~\eqref{Ddd} is simplified as
$\Ddd=\dD$, and the Fourier kernel for the monopole power spectrum is
\beeq
\TTT^V_0(k,\tilde k)=\sqrt{2\over\pi}k
\int d\rbar~\rbar^2j_0(k\rbar){j_0'(\tilde k\rbar)\over \tilde k\rbar}
(\HH\rbar-1)f\TT_m(\tilde k;\rbar)~,
\eneq
where we used Eq.~\eqref{single} and there is no contribution
of~$\kappa$ or~$V_0$ to the monopole power. While the radial correlation
is properly considered, the dominant source (or $\de_g$) of the correlation
is missing, and this assumption still {\it underestimates} 
the minimum errors.

Two symbols in Figure~\ref{fig:error} show the current errors 
from the Pantheon sample \cite{PSSN18} with $\zm\simeq2.3$
and the Supernova Legacy Survey (SNLS; \cite{SNLS14}) with $\zm\simeq1.0$.
The errors from both surveys in Figure~\ref{fig:error}
are those reported in the collaboration papers \cite{SNLS14,PSSN18}, 
in which they combined the supernova observations with the
 {\it Planck} CMB analysis. The intrinsic scatter associated
with individual supernovae is a dominant contribution to their error budget,
which we set zero in idealized surveys considered here. Moreover,
while the correlation of the peculiar velocities is approximately accounted 
for in the analysis \cite{SNLS14,PSSN18}, 
the correlation of the host galaxies is ignored.
While the fundamental limit set by the cosmic variance is yet to be reached
in these surveys
at high redshift, its impact is more significant at lower redshift.

\section{Discussion}
Accounting for the correct correlation of the 
luminosity distance measurements, we derived the {\it precise}
minimum errors on two 
cosmological parameters from idealized supernova surveys, where an infinite 
number of supernova observations are made without any systematic errors. 
The minimum floors exist because the light propagation from individual 
supernovae is affected by large-scale inhomogeneities and the host galaxies
are also correlated. However, past work in literature ignored the radial
correlation along the past light cone or the contribution of the supernova
host galaxies. With all the correlations properly accounted for, we find that
the luminosity distance measurements are more correlated and the cosmological
constraining power is more reduced than previously estimated.

Today, the systematic errors associated with the light-curve calibration 
are the limiting factors in the current supernova surveys, and they cannot
be beaten down by increasing the number of observations in future surveys.
Our results from idealized supernova surveys without such systematic errors,
however, provide fundamental limitations set by the cosmic variance, 
regardless of the survey specifications in the future. In practice, one
would have to consider both contributions
in designing a supernova survey. Apparent from
Figure~\ref{fig:error}, the cosmic variance is more important at low
redshift, where only a small fraction of the Universe can be probed.

Though we assumed perfect knowledge of 
$\Delta^2_\varphi$, $\TT_m$, $b$, $\alpha$, and~$f$ for the computation,
their uncertainties in practice would also degrade the cosmological 
constraining power. In particular, the bias factor for the supernova
host galaxies is expected to be different from that of typical galaxies,
and its theoretical modeling is highly uncertain.
While the second-order fluctuations from inhomogeneities will break our
assumption of Gaussianity, its impact is largely limited to shifting the
best-fit cosmological parameters at a percent level \cite{BEDUET14},
and hence we believe that the impact on the cosmic variance is similar 
in magnitude. While the real surveys are yet to reach the limits 
considered here, there are several ways to extract more information and
overcome the limits in Figure~\ref{fig:error}. The spatial 
correlation~$\xi_{ij}$ of 
individual observations contains cosmological information that can be
harnessed, but was ignored in Figure~\ref{fig:error}. Cross correlation
with other galaxy populations in the same volume provides a way to beat
the cosmic variance and reduce the errors \cite{SELJA09}.

\acknowledgments
We thank Ermis Mitsou for useful discussions. We acknowledge support
by the Swiss National Science Foundation. J.Y. is further supported by
a Consolidator Grant of the European Research Council (ERC-2015-CoG grant 
680886).

\bibliography{ms.bbl}

\end{document}